\documentclass[12pt,preprint]{aastex}

\newcounter{address}
\newcommand{\latin}[1]{{#1}}
\newcommand{\ie}{\latin{i.e.}}
\newcommand{\eg}{\latin{e.g.}}
\newcommand{\cf}{\latin{c.f.}}

\newcommand{\vs}{\latin{vs.}}
\newcommand{\Halpha}{\ensuremath{\mathrm{H}\alpha}}
\newcommand{\Hbeta}{\ensuremath{\mathrm{H}\beta}}
\newcommand{\Hgamma}{\ensuremath{\mathrm{H}\gamma}}
\newcommand{\Hdelta}{\ensuremath{\mathrm{H}\delta}}
\newcommand{\OII}{\ensuremath{\mathrm{[O\,II]}}}
\newcommand{\OIII}{\ensuremath{\mathrm{[O\,III]}}}
\newcommand{\Kamp}{\ensuremath{\textsf{K}}}
\newcommand{\Aamp}{\ensuremath{\textsf{A}}}

\begin{document}
\title{
  Selection and photometric properties of K+A galaxies
}
\author{
  Alejandro~D.~Quintero\altaffilmark{\ref{NYU}},
  David~W.~Hogg\altaffilmark{\ref{NYU},\ref{email}},
  Michael~R.~Blanton\altaffilmark{\ref{NYU}},
  David~J.~Schlegel\altaffilmark{\ref{Princeton}},
  Daniel~J.~Eisenstein\altaffilmark{\ref{Steward}},
  James~E.~Gunn\altaffilmark{\ref{Princeton}},
  J.~Brinkmann\altaffilmark{\ref{APO}},
  Masataka~Fukugita\altaffilmark{\ref{Tokyo}},
  Karl~Glazebrook\altaffilmark{\ref{JHU}},
  and
  Tomotsugu~Goto\altaffilmark{\ref{Tokyo},\ref{CMU}}
}

\setcounter{address}{1}
\altaffiltext{\theaddress}{\stepcounter{address}\label{NYU}
Center for Cosmology and Particle Physics, Department of Physics, New
York University, 4 Washington Place, New York, NY 10003}
\altaffiltext{\theaddress}{\stepcounter{address}\label{email}
To whom correspondence should be addressed: \texttt{david.hogg@nyu.edu}}
\altaffiltext{\theaddress}{\stepcounter{address}\label{Princeton}
Princeton University Observatory, Princeton, NJ 08544}
\altaffiltext{\theaddress}{\stepcounter{address}\label{Steward}
Steward Observatory, 933 N. Cherry Ave., Tucson, AZ 85721}
\altaffiltext{\theaddress}{\stepcounter{address}\label{APO}
Apache Point Observatory, 2001 Apache Point Road,
P.O. Box 59, Sunspot, NM 88349}
\altaffiltext{\theaddress}{\stepcounter{address}\label{Tokyo}
Institute for Cosmic Ray Research, University of Tokyo,
Kashiwa, 2778582, Japan}
\altaffiltext{\theaddress}{\stepcounter{address}\label{JHU}
Department of Physics and Astronomy, The Johns Hopkins University,
Baltimore, MD 21218}
\altaffiltext{\theaddress}{\stepcounter{address}\label{CMU}
Department of Physics, Carnegie Mellon University,
5000 Forbes Avenue, Pittsburgh, PA 15213}

\begin{abstract}
Two different simple measurements of galaxy star formation rate with
different timescales are compared empirically on $156,395$ fiber
spectra of galaxies with $r<17.77$~mag taken from the Sloan Digital
Sky Survey in the redshift range $0.05<z<0.20$: a ratio $\Aamp /
\Kamp$ found by fitting a linear sum of an average old stellar
poplulation spectrum (\Kamp) and average A-star spectrum (\Aamp) to
the galaxy spectrum, and the equivalent width (EW) of the $\Halpha$
emission line.  The two measures are strongly correlated, but there is
a small clearly separated population of outliers from the median
correlation that display excess $\Aamp /\Kamp$ relative to \Halpha EW.
These ``K+A'' (or ``E+A'') galaxies must have dramatically decreased
their star-formation rates over the last $\sim 1$ Gyr.  The K+A
luminosity distribution is very similar to that of the total galaxy
population.  The K+A population appears to be bulge-dominated, but
bluer and higher surface-brightness than normal bulge-dominated
galaxies; it appears that K+A galaxies will fade with time into normal
bulge-dominated galaxies.  The inferred rate density for K+A galaxy
formation is $\sim 10^{-4}\,h^3~\mathrm{Mpc^{-3}\,Gyr^{-1}}$ at
redshift $z\sim 0.1$.  These events are taking place in the field; K+A
galaxies don't primarily lie in the high-density environments or
clusters typical of bulge-dominated populations.
\end{abstract}


\section{Introduction}

Cosmic star formation activity appears to be coming to an end.  All
indicators of star formation show that the cosmic mean comoving
density of star formation has been declining since redshifts near
unity \citep[\eg,][and references therein]{hogg01sfr}.  The bulk of
stars in the local Universe are found in old stellar populations
\citep{fukugita98a, hogg02a}.  One consequence of this global decline
is that galaxies in the local Universe ought to be shutting down their
individual star formation activities.  It is not understood whether
the changes in star formation in individual galaxies are expected to
be gradual or dramatic.

Our extremely local neighborhood in the disk of the Milky Way has been
forming stars at a relatively constant rate for the last few billion
years \citep{barry88a,pardi94,prantzos98,rocha-pinto00,gizis02a}, but
it appears that it can't continue for much longer. When the total
available gas content is compared to the star formation rate, it
appears that the Milky Way will run out of gas in the next few Gyr
\citep[\eg,][]{mooney88a, blitz96a}.  This is also true for the
majority of local disk galaxies \citep{kennicutt94a}. In detail these
calculations do not account for gas recycling, gas flux from plasma
resevoirs, or new gas accretion, but the trend they imply for cosmic
star formation activity is consistent with that found in cosmological
observations.

Here we begin a comparison of different optical star-formation rate
measurements in a large sample of galaxies with spectroscopic
observations in the Sloan Digital Sky Survey \citep[SDSS;
\eg,][]{york00a}.  One star-formation rate measurement makes use of
the strength of the \Halpha\ emission line, which is emitted in
ionization regions around O and B stars, with lifetimes of $\sim
10^7~\mathrm{yr}$.  Another uses the fraction of the galaxy light
emitted by A stars, with lifetimes of $\sim 10^9~\mathrm{yr}$.  These
star-formation indicators are sensitive to stars of different masses
and lifetimes.

Star-formation indicators tracking stars with different lifetimes can
be compared to flag galaxies with strong star-formation rate
derivatives.  In particular, comparison of our \Halpha\ and A-star
measurements locates galaxies that have changed their star-formation
rates within the last $\sim 10^9~\mathrm{yr}$.  If star formation is
triggered by mergers or interactions, such galaxies might be found to
be undergoing or recovering from an interaction, and their statistics
might provide a measure of the merger rate.  If star-formation
efficiency is a strong function of local environment, then galaxies
with negative star-formation derivatives might be tracers of the
environmental conditions at the transition between those that are
conducive to star-formation and those that aren't.

One population of galaxies with strong negative star-formation rate
derivatives is already known: ``K+A'' or ``E+A'' galaxies
\citep{dressler83a, zabludoff96a, goto03a}.  These are galaxies
combining the stellar absorption lines of A stars with those of old
stellar populations, but, at the same time, showing little sign of
current star-formation.  K+A galaxies are exceedingly rare---they
comprise between $10^{-2}$ and $\sim 10^{-4}$ of the galaxy population
(depending on definition)---but they are expected to provide important
information about the conditions under which stars form and galaxies
evolve.

For example, if it is true that elliptical or bulge-dominated galaxies
are formed by the mergers of spiral or disk-dominated galaxies, then
the central concentration of stars, the alpha-enhanced chemical
abundances \citep[\eg,][]{worthey98a}, and the lack of cold gas, dust
and star formation \citep[\eg,][]{roberts94a} all require large, brief,
centrally concentrated starbursts during or immediately following
disk--disk mergers.  Note that the abundance ratios provide the
evidence that the starbursts must be brief.  These merger events
therefore likely go through a K+A phase for $\sim 10^9~\mathrm{yr}$;
in this scenario, it should be possible to use the abundances of K+A
galaxies to constrain merger rates.  Indeed, it has already been shown
that the A stars tend to be centrally concentrated in K+A galaxies and
that some have ``kinematically hot'' stellar populations like
bulge-dominated galaxies \citep{norton01a}.  On the other hand, in
some K+A galaxies, there is residual gas, implying that the star
formation event does not always exhaust the entire gas reservoir
\citep{Chang01a}.

Although any scenario in which bulge-dominated galaxies are created by
mergers virtually requires the remnants to be K+A events, an apparent
excess of A stars over O and B stars can be produced in heavily
dust-enshrouded star-formation regions from which young stars escape
on Gyr timescales, or in mergers in which an old galaxy captures a
smaller, younger galaxy, shutting off star formation in the process.

In what follows, a cosmological world model with
$(\Omega_\mathrm{M},\Omega_\mathrm{\Lambda})=(0.3,0.7)$ is adopted,
and the Hubble constant is parameterized
$H_0=100\,h~\mathrm{km\,s^{-1}\,Mpc^{-1}}$, for the purposes of
calculating distances and volumes \citep[\eg,][]{hogg99cosm}.

\section{Data}

The SDSS is taking $ugriz$ CCD imaging of $10^4~\mathrm{deg^2}$ of the
Northern Galactic sky, and, from that imaging, selecting $10^6$
targets for spectroscopy, most of them galaxies with
$r<17.77~\mathrm{mag}$ \citep[\eg,][]{gunn98a,york00a,stoughton02a}.

All the data processing, including astrometry \citep{pier03a}, source
identification, deblending and photometry \citep{lupton01a},
calibration \citep{fukugita96a,smith02a}, spectroscopic target
selection \citep{eisenstein01a,strauss02a,richards02a}, and
spectroscopic fiber placement \citep{blanton03a} are performed with
automated SDSS software.

Every spectral ``plate'' of fiber positions includes several faint
($15.5$ to $18.5$~mag) F stars.  The spectra are calibrated with these
F-star spectra; ie, they are multiplied by the function of wavelength
that makes the F-star spectra match F-star spectrophotometry.
Although this calibration procedure produces consistent calibration at
the 10-percent level \citep[as measured by comparisons of old stellar
populations at various redshifts,][]{eisenstein03b}, it does not deal
carefully with the diversity of F-star spectra, or the possibility
that some of the F stars, even at inferred distances $>2$~kpc, may be
inside some part of the Galactic extinction.  Inasmuch as the
calibration stars are beyond the Galactic extinction, however, the
spectra are effectively extinction-corrected.

Redshifts are measured on the reduced spectra by an automated system,
which models each galaxy spectrum as a linear combination of stellar
eigenspectra (Schlegel, in preparation).  The eigenspectra are
constructed over the rest-frame wavelength range
$4100<\lambda<6800$~\AA\  from a high-resolution spectroscopic library
\citep{prugniel01a}.  The central velocity dispersion $\sigma_v$ is
determined by fitting the detailed spectral shape as a
velocity-smoothed sum of stellar spectra (Schlegel \& Finkbeiner, in
preparation).

It must be emphasized that the 3~arcsec diameter spectroscopic fibers
of the SDSS spectrographs do not obtain all of each galaxy's light
because at redshifts of $0.05<z<0.2$ they reprsent apertures of
between $2$ and $7~h^{-1}\,\mathrm{kpc}$ diameter.  For this reason,
in what follows, SDSS imaging data rather than spectroscopy are used
to infer the global properties of the galaxies.

Galaxy luminosities and colors are computed in fixed bandpasses, using
Galactic extinction corrections \citep{schlegel98a} and $K$
corrections \citep[computed with \texttt{kcorrect
v1\_11};][]{blanton03b}.  They are $K$ corrected not to the redshift
$z=0$ observed bandpasses but to bluer bandpasses $^{0.1}g$, $^{0.1}r$
and $^{0.1}i$ ``made'' by shifting the SDSS $g$, $r$, and $i$
bandpasses to shorter wavelengths by a factor of 1.1
\citep[\cf,][]{blanton03b, blanton03d}.  This means that galaxies at
redshift $z=0.1$ (typical of the sample used here) have trivial
$K$ corrections.

To the azimuthally averaged radial profile of every galaxy in the
observed-frame $i$ band, a seeing-convolved S\'ersic model is fit, as
described elsewhere \citep{blanton03d}.  The S\'ersic model has
surface brightness $I$ related to angular radius $r$ by $I\propto
\exp[-(r/r_0)^{(1/n)}]$, so the parameter $n$ (S\'ersic index) is a
measure of radial concentration (seeing-corrected).  At $n=1$ the
profile is exponential, and at $n=4$ the profile is de~Vaucouleurs.

To every best-fit S\'ersic profile, the \citet{petrosian76a}
photometry technique is applied, with the same parameters as used in
the SDSS survey.  This supplies seeing-corrected Petrosian magnitudes
and radii.  A $K$-corrected surface-brightness $\mu_{^{0.1}i}$ in the
$^{0.1}i$ band is computed by dividing half the $K$-corrected
Petrosian light by the area of the Petrosian half-light circle.

For the purposes of computing large-scale structure statistics, we
have assembled a complete subsample of SDSS galaxies known as the NYU
LSS \texttt{sample12}.  This subsample is described elsewhere
\citep[\eg,][]{blanton03d}; it is selected to have a well-defined
window function and magnitude limit.  In addition, the sample of
galaxies used here was selected to have apparent magnitude in the
range $14.5<r<17.77~\mathrm{mag}$, redshift in the range
$0.05<z<0.20$, and absolute magnitude $M_{^{0.1}i}>-24~\mathrm{mag}$.
These cuts left 156,395 galaxies.

For each galaxy, a selection volume $V_\mathrm{max}$ is computed,
representing the total volume of the Universe (in
$h^{-3}~\mathrm{Mpc^3}$) in which the galaxy could have resided and
still made it into the sample.  The calculation of these volumes is
described elsewhere \citep{blanton03c, blanton03d}.  For each galaxy,
the quantity $1/V_\mathrm{max}$ is that galaxy's contribution to the
cosmic number density, and the quantity $L/V_\mathrm{max}$, where $L$
is the luminosity in some bandpass, is that galaxy's contribution to
the luminosity density in that bandpass.

Around each galaxy in a complete subsample, \texttt{sample10}, of
\texttt{sample12}, there is a measure of the overdensity $\delta_1$
inside a spherical Gaussian window of radius $1~h^{-1}\,\mathrm{Mpc}$,
made by deprojecting the angular distribution of nearby galaxies
detected in the imaging at fluxes corresponding to luminosities within
1~mag of $L^\ast$ at the target galaxy's redshift, as described
elsewhere \citep{eisenstein03a, hogg03b}.  The individual overdensity
estimates are low in signal-to-noise, but they are unbiased in the
mean when averaged over sets of galaxies.  A galaxy in an environment
with the cosmic mean density has $\delta_1=0$.

There is another measure $\delta_8$ on an $8\,h^{-1}~\mathrm{Mpc}$
scale, that is a measure of the three-dimensional comoving number
density excess around each galaxy in a sphere of $8\,h^{-1}$~Mpc
comoving radius, with no correction for peculiar velocities, as
described elsewhere \citep{hogg03b, blanton03d}.  Neighbor galaxies
are counted, the result is divided by the prediction made from the
galaxy luminosity function \citep{blanton03c}, and unity is subtracted
to produce the overdensity estimate.  The estimates $\delta_8$ are
higher in signal-to-noise than the estimates $\delta_1$, but they are
slightly biased because they do not account for peculiar velocities.

\section{Method}

An average old stellar population spectrum (hereafter called ``K''
because it is dominated by K-type stars) was made by averaging the
spectra of bulge-dominated galaxies with luminosities near $L^{\ast}$
and in typical environments, as described previously
\citep{eisenstein03b}.  An average A-star spectrum (``A'') was
obtained by scaling and taking the mean of calibrated SDSS spectra of
stars with types determined by eye to be near A0. These two spectra
were normalized by scaling them to have equal total flux within the
wavelength range $3800<\lambda<5400$~\AA.


A linear sum of the K and A spectra was fit to every SDSS spectrum in
the wavelength range $3800<\lambda<5400$~\AA, weighting by the inverse
square uncertainty in the observed spectrum.  A
$280~\mathrm{km\,s^{-1}}$ region was masked out around \Hbeta\ and the
\OIII\ lines; this mask size was determined by observing that it would
mask out the great majority of the emission-line flux in the great
majority of spectra.  Before fitting, the A spectrum was smoothed to
the appropriate velocity dispersion for each spectrum.
\cite{dressler83a} pioneered this method by showing that
post-starburst galaxies can be modeled well by an A dwarf spectrum and
a K0 spectrum over the range of 3400 to 5400 angstroms. They found, as
we do, that this model simultaneously describes the broad spectral
continuum and the depths of narrow features.

The fitting procedure returns two numbers, the amplitudes \Kamp\ and
\Aamp\ of the K and A spectral components in the best fit.

The \Halpha\ line flux is measured in a 20~\AA\ width interval centered
on the line.  Before the flux is computed, the best-fit model A+K
spectrum is scaled to have the same flux continuum as the data in the
vicinity of the emission line and subtracted to leave a
continuum-subtracted line spectrum.  This method fairly accurately
models the \Halpha\ absorption trough in the continuum, although in
detail it leaves small negative residuals, as will be shown below.
The flux is converted to a rest-frame equivalent width (EW) with a
continuum found by taking the inverse-variance-weighted average of two
sections of the spectrum about 150~\AA\ in size and on either side of
the emission line.

Some general caveats apply to our measurements.  The first is that the
K+A fit is not always good; in particular, galaxies that have
undergone a very recent ($<10^{8}~\mathrm{yr}$) starburst can have the
blue end of their spectra dominated by O and B stars, not A stars.
These galaxies will be assigned high $\Aamp /\Kamp$, because the A
template fits better than the K, even though the A stars are not
dominating the spectra.  Of course, most of these galaxies also show
strong nebular \Halpha\ emission, and, indeed, they do contain young
stars.

We rely heavily on the flux-calibration of the spectra, because $\Aamp
/\Kamp$ is sensitive to the color or tilt of the spectrum.  If there
were large ($>20~\mathrm{percent}$) absolute calibration deviations
among the galaxy spectra, we might introduce some spurious
interlopers.  However, the peak of the $\Aamp /\Kamp$ distribution of
galaxies with low \Halpha EW at $\Aamp /\Kamp =0$ is narrow and
symmetrical about zero.  This is a strong indicator that the
spectrophotometric calibration is well-behaved.  Note that we rely on
the consistency of SDSS spectrophotometric calibration, but not on its
absolute accuracy.  The typical standard linear fitting errors in
$\Aamp /\Kamp$ are 0.03 to 0.1, roughly consistent with the scatter
around $\Aamp /\Kamp =0$.

We are looking at fiber spectra, inside the central 3~arcsec diameter
of each galaxy.  The physical size of this aperture is between $2$ and
$7~h^{-1}\,\mathrm{kpc}$ in diameter, thereby excluding light from the
outer parts of the galaxies.  Some of our galaxies classified as K+A
could, in principle, have active star formation in their outskirts.
Strictly speaking, the K+A classification is a classification of the
star formation indicators in the inner regions of galaxies.  For this
reason, in what follows the SDSS imaging rather than spectroscopy is
used to measure global properties of the galaxies.

The fitting procedure doesn't take dust exctinction into account.  The
$\Aamp /\Kamp$ ratio is sensitive to the shape of the blue end of the
spectrum, so it will be reduced in a galaxy in which the A stars are
behind significant amounts of dust.  This is also true of the \Halpha
EW measurements, although less so because \Halpha\ is redward of the
K+A fitting region.  For these reasons, there may be some
truly post-starburst galaxies not classified as K+A galaxies because
their young stars are dust-enshrouded.

The K+A population defined by this project does not contain any
significant AGN because the selection permits so little \Halpha\
emission.  There may be some truly post-starburst galaxies not
classified as K+A galaxies because they contain central AGN and
therefore emit significant \Halpha\ emission.

\section{Model spectra}

The \Kamp, \Aamp, and \Halpha\ measurements were made also on model
spectra made by convolving PEGASE \citep{fioc97a} instantaneous
starburst models with different star formation histories.

The original instantaneous starburst history was made with solar
metallicity.  Although nebular emission is included in the model,
there was no modeling of galactic winds, infall, or internal
extinction.  For each timestep, the \Kamp\ and \Aamp\ measurements
were computed with exactly the same least-square-fitting procedure as
was performed on the data. The continuum around the \Halpha\ line was
also computed in exactly the same way as in the data, but the flux of
the \Halpha\ line was taken directly from the PEGASE outputs, because
PEGASE returns integrated flux values for the emission lines rather
than flux densities.

Three model spectra and associated evolutionary tracks were computed
by convolving the instantaneous PEGASE burst model: A galaxy forming
stars at a fixed rate for 14~Gyr; a galaxy forming stars at a fixed
rate for 10~Gyr, shutting off, and fading quiescently for 4~Gyr; and a
galaxy forming stars at a fixed rate for 3~Gyr, shutting off, and
fading quiescently for 11~Gyr.

In the end, because of small differences between the old model spectra
and old observed galaxies, the final \Kamp\ and \Aamp\ values were
transformed (by a linear transformation equivalent to a
flux-conserving shear in the \Kamp\ \vs\ \Aamp\ plane) so that an old
model galaxy shows $\Aamp /\Kamp=0$, just like an old real galaxy.
The linear transformation affects the model outputs by less than
15~percent.

\begin{figure}
\plotone{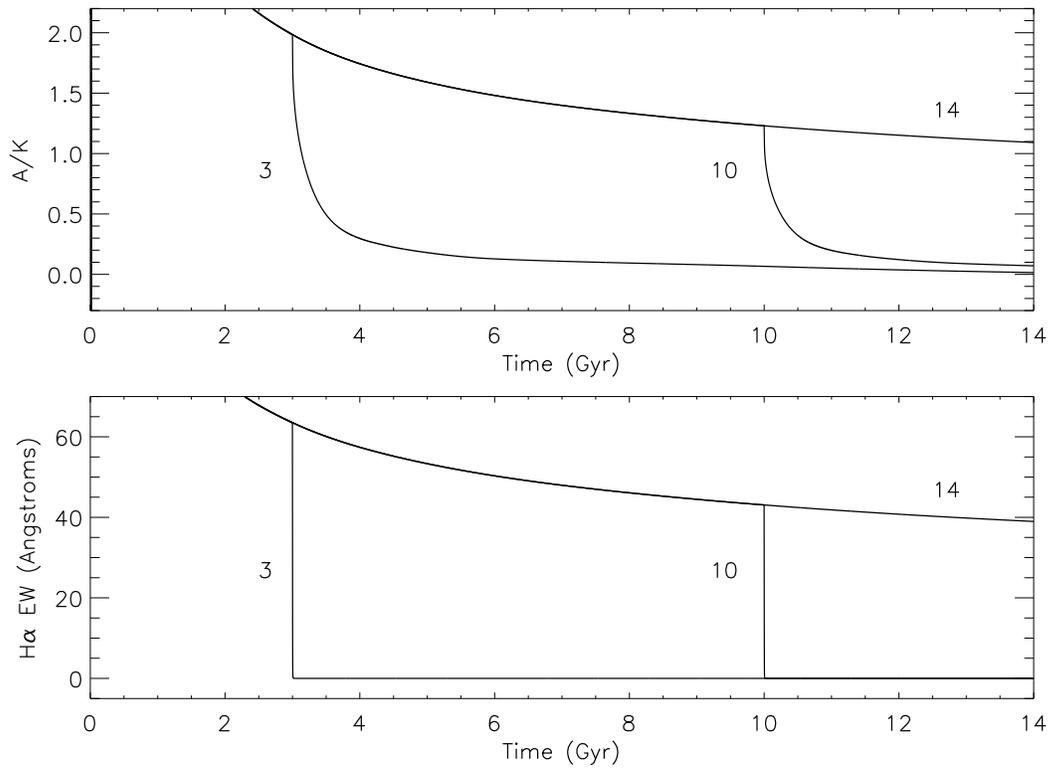}
\caption{The indicators $\Aamp /\Kamp$ and \Halpha\ plotted against
time for the three star formation models with constant star formation
for 14, 10, and 3~Gyr, followed by quiescent fading (see text for
details).\label{fig:Poster_timemodels}}
\end{figure}

The measurements $\Aamp / \Kamp$ and \Halpha EW are plotted as a
function of cosmic time for the three models in
Figure~\ref{fig:Poster_timemodels}.


\section{Results}

The measured $\Aamp /\Kamp$ values for the galaxies in the sample are
plotted against the measured \Halpha EW values in
Figure~\ref{fig:Poster_AE_vs_Halpha}.  The two star-formation
indicators are highly correlated, as expected.  Overplotted on the
figure are models of galaxies with star formation histories consisting
of constant star-formation rates followed by periods of total
quiescence.

It is worthy of note that most of the galaxies appear more quiescent
(\ie, have lower $\Aamp /\Kamp$ and lower \Halpha EW) than the model
(labeled ``14'') in which stars form at a constant rate over the
entire lifetime of the Universe.  In other words, the present-day
star-formation rate appears much lower than the cosmic time-averaged
mean.

The models in which star formation has ``shut off'' drop precipitously
in \Halpha EW (which indicates stars with lifetimes of $\sim 10^7$~yr)
and then slowly in $\Aamp /\Kamp$ (which indicates stars with
lifetimes of $\sim 10^9$~yr.  The models suggest that galaxies whose
star-formation activities have shut off will evolve into the
zero-star-formation population along the zero \Halpha EW line.
Indeed, Figure~\ref{fig:Poster_AE_vs_Halpha} shows a clear ``spur'' of
galaxies along this line.  These are the ``K+A galaxies''.

\begin{figure}
\plotone{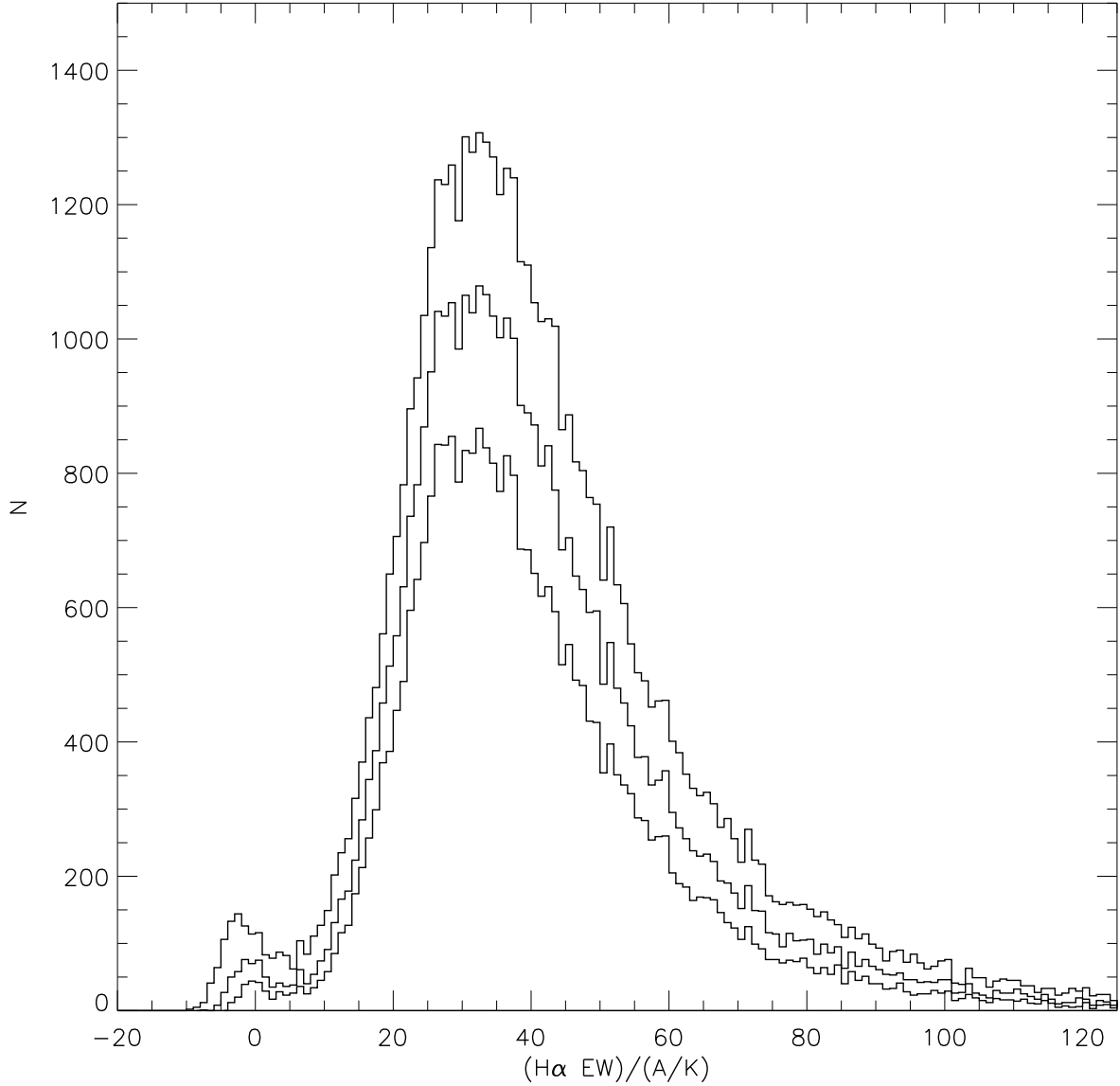}
\caption{The distributions of the ratio of the two star formation
indicators, \Halpha\ EW and $\Aamp /\Kamp$, for four subsamples of
$\texttt{sample12}$, selected to have $\Aamp /\Kamp > 0.2$, 0.3,
and 0.4.  Note the evidence for the existence of a separate population
of K+A galaxies around zero.\label{fig:Poster_histograms}}
\end{figure}

Figure~\ref{fig:Poster_histograms} shows the ratio of \Halpha EW to
$\Aamp / \Kamp$ for samples cut in $\Aamp / \Kamp$.  The histograms
are bimodal, showing a population with anomalously low \Halpha EW.
These are the K+A galaxeis.  The figure shows that K+A galaxies form
an identifiable separate population. This figure also suggests that
the line that separates the ``K+A galaxies'' from the rest of the
population has a slope of $(\Halpha\mathrm{EW})/(\Aamp / \Kamp)=5$ on
Figure~\ref{fig:Poster_AE_vs_Halpha}. This figure also shows that by
defining ``K+A galaxies'' to have $\Aamp /\Kamp > 0.2$ we are
excluding a large number of transition objects.  The cut was put at
0.2 to avoid interlopers with true values of $\Aamp /\Kamp \approx
0.0$ but scattered high by calibration or statistical errors.

\begin{figure}
\plotone{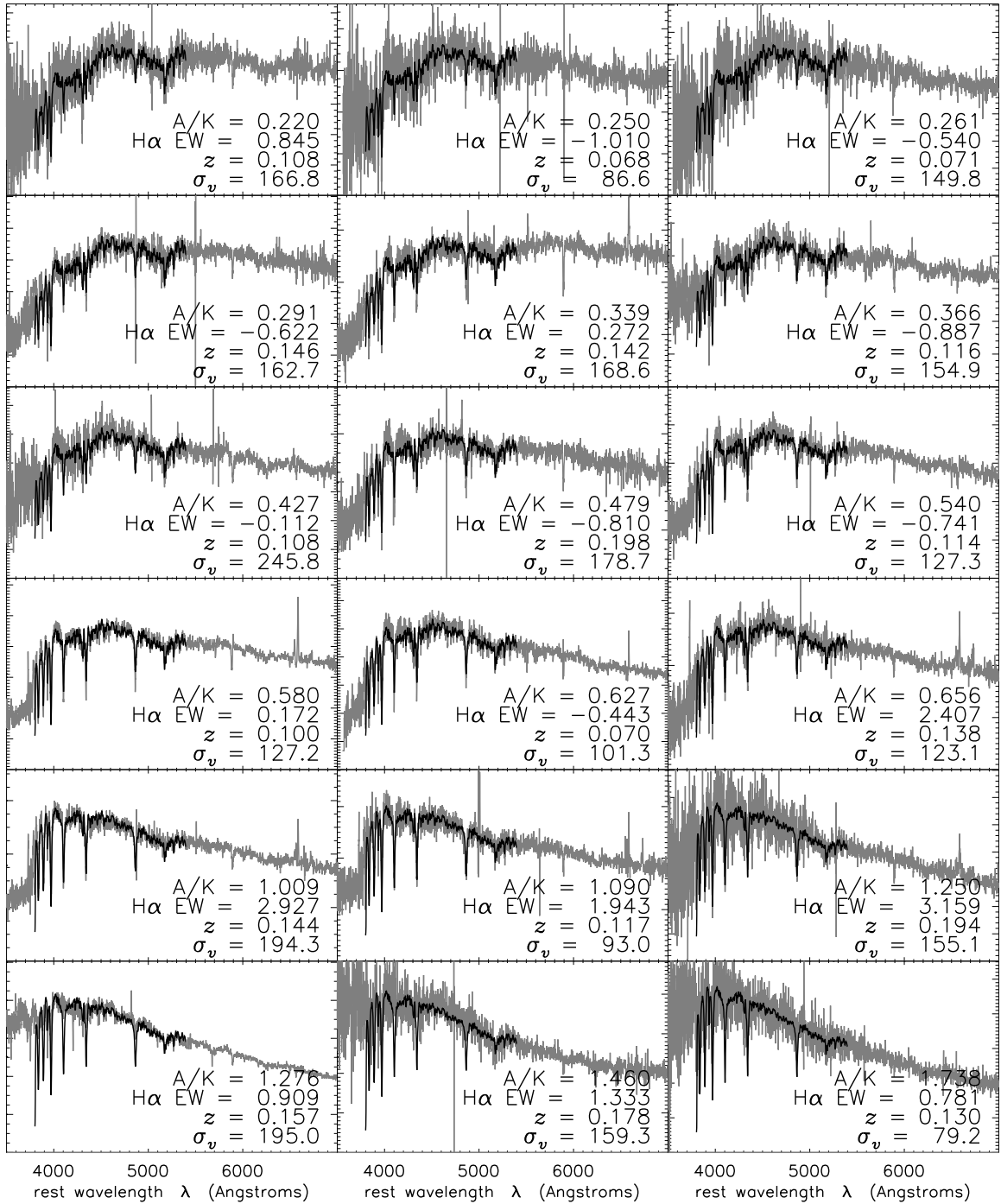}
\caption{Spectra of members of a randomly selected subsample of the
K+A population, selected from $\Aamp /\Kamp$ bins so as to span the
range. The grey lines show the data and the black overplotted lines
show the ``K+A'' best fit.\label{fig:Poster_catalog}}
\end{figure}

The selection of K+A galaxies for what follows is indicated by the
green lines on Figure~\ref{fig:Poster_AE_vs_Halpha}.  There are 1194
K+A galaxies within these selection boundaries in
\texttt{sample12}. Spectra of a randomly selected subsample of this
population is shown in Figure~\ref{fig:Poster_catalog}.

The astute reader will notice that the mode \Halpha EW is less than
zero; this offset is due to the crudeness of the continuum model
subtracted during \Halpha\ line measurement.  This offset does not
affect any of our conclusions.

It is interesting to compare our selection criteria with those of
\citet{zabludoff96a} and \citet{goto03a}.  \citet{zabludoff96a}
considered galaxies in the redshift range $0.05<z<0.13$ with \OII\
emission $EW<2.5$~\AA\ and \Hgamma, \Hdelta, and \Hbeta\ absorption
$EW>5.5$~\AA. They found 21 K+A galaxies out of a sample of
11113 (roughly 0.2~percent).  \citet{goto03a} also worked with an SDSS
sample of galaxies, but in the redshift range $0.05<z<1.0$ and with
$\Hdelta>4$~\AA.  They found 3340 HDS galaxies out of a sample of
95479 (roughly 3.4~percent).

\begin{figure}
\plotone{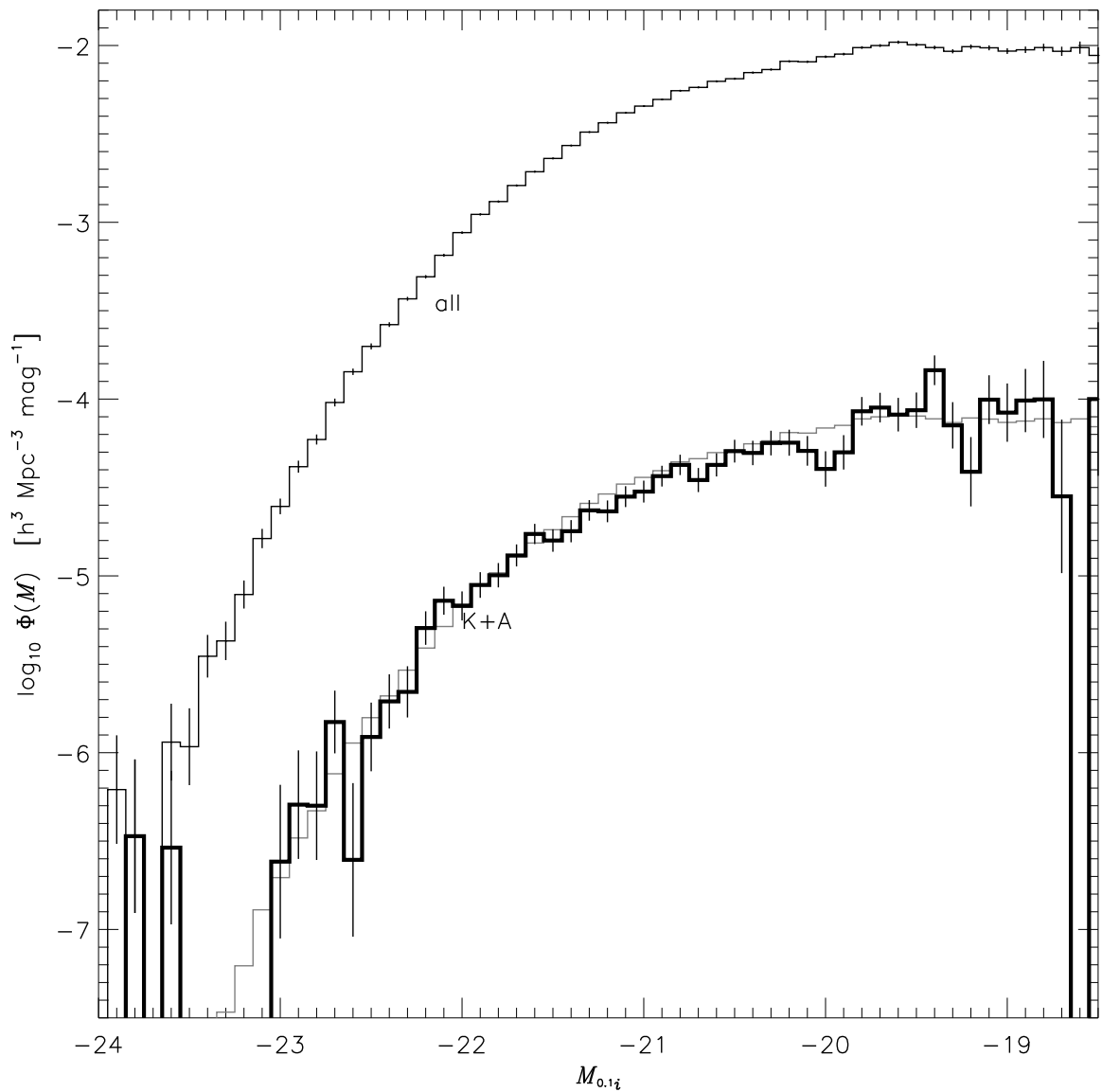}
\caption{The luminosity function---\ie, luminosity distribution of the
sample weighted by $1/V_\mathrm{max}$---for the entire sample and the
K+A galaxies.  The plotted uncertainties are simply the
$1/V_\mathrm{max}$ weights added in quadrature (divided by the bin
width).\label{fig:Poster_lumweight}}
\end{figure}

Figure~\ref{fig:Poster_lumweight} shows the luminosity function of the
K+A galaxies compared with the luminosity function of the entirety of
\texttt{sample12}.  The shapes are very similar.  Interestingly, the
shape of the K+A luminosity function appears closer to the shape of
the total luminosity function than it does to the early-type galaxy or
red galaxy luminosity functions \citep[\eg,][]{blanton03c}.  In
detail, quantitative comparisons of the luminosity functions ought to
involve the fading with time of the K+A galaxy light.

Figure~\ref{fig:Poster_aehist} shows the distribution of $\Aamp
/\Kamp$ for the K+A galaxies, in number-density units (the ``ratio
function''?).  Overplotted are the expected distributions of $\Aamp
/\Kamp$ under the assumption of steady-state creation of K+A galaxies
along evolutionary tracks like those of the ``10'' model (\ie, 10~Gyr
of constant star formation followed by nothing) or the ``3'' model
(\ie, 3~Gyr of constant star formation followed by nothing).  Neither
of these models is a good description, suggesting that there is
probably a large diversity to the star-formation histories that end in
a K+A phase according to the selection criteria used here.

\begin{figure}
\plotone{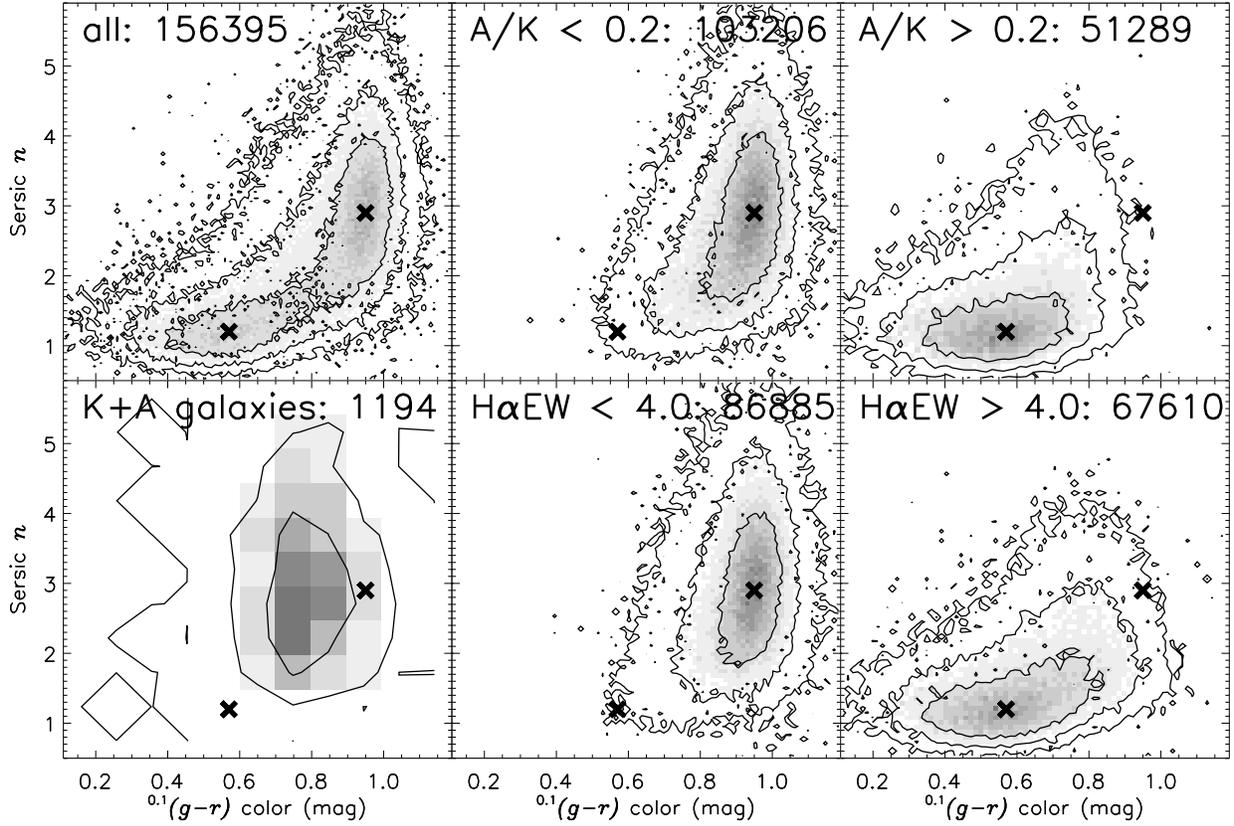}
\caption{The top left panel shows the distribution of luminosity
density in galaxies in the sample in the color--S\'ersic-index plane.
The color is measured in blueshifted SDSS bandpasses $^{0.1}g$ and
$^{0.1}r$, and the S\'ersic index $n$ is found by fitting to the
radial profile in the $^{0.1}i$ band.  Each galaxy data point has been
weighted by the ratio of its luminosity $L_{^{0.1}i}$ in the $^{0.1}i$
band to its comoving selection volume $V_\mathrm{max}$, so the
distribution of weight in the plot is proportional to comoving
luminosity density.  The contours enclose 52.0, 84.3, and 96.6~percent
of the total luminosity density.  In this panel, galaxies separate
into disk-dominated (blue, low $n$) and bulge-dominated (red, high
$n$) populations, marked by crosses.  The other panels show similar
plots, but for different sub-samples, cut on $\Aamp /\Kamp$ and
\Halpha EW.  The ``K+A'' panel shows those galaxies making the high
$\Aamp /\Kamp$ cut and the low \Halpha EW cut.  Galaxies selected to
have low $\Aamp /\Kamp$ or low \Halpha EW appear to be concentrated
(bulge-dominated); galaxies selected to have high $\Aamp /\Kamp$ or
high \Halpha EW appear to be exponential (disk-dominated).  The K+A
galaxies have the colors of disk-dominated galaxies but many have the
radial profiles of bulge-dominated galaxies.\label{fig:manyd_gmr_n}}
\end{figure}

Figure~\ref{fig:manyd_gmr_n} shows the luminosity-density weighted
distribution of all galaxies in the plane of $^{0.1}(g-r)$ color and
S\'ersic index $n$ (seeing-corrected concentration).  As discussed
elsewhere \citep{blanton03d}, the general galaxy population is bimodal
in this plane, showing a population of blue, exponential
(disk-dominated) galaxies and a separate population of red,
concentrated (bulge-dominated) galaxies.
Figure~\ref{fig:manyd_gmr_n} also shows that galaxies chosen to have
high star-formation rates, by either $\Aamp /\Kamp$ or
\Halpha EW, lie predominantly in the blue, exponential class.
Galaxies chosen to have low star-formation rates lie predominantly in
the red, concentrated class.  Galaxies chosen to have high
$\Aamp /\Kamp$ but low \Halpha EW---the K+A galaxies---have
some members clearly in the blue, exponential class, and some in an
outlier class that appears concentrated, but blue.  If star formation
has truly been shut off in these galaxies, they are expected to fade
and redden and become part of the red, concentrated class.

\begin{figure}
\plotone{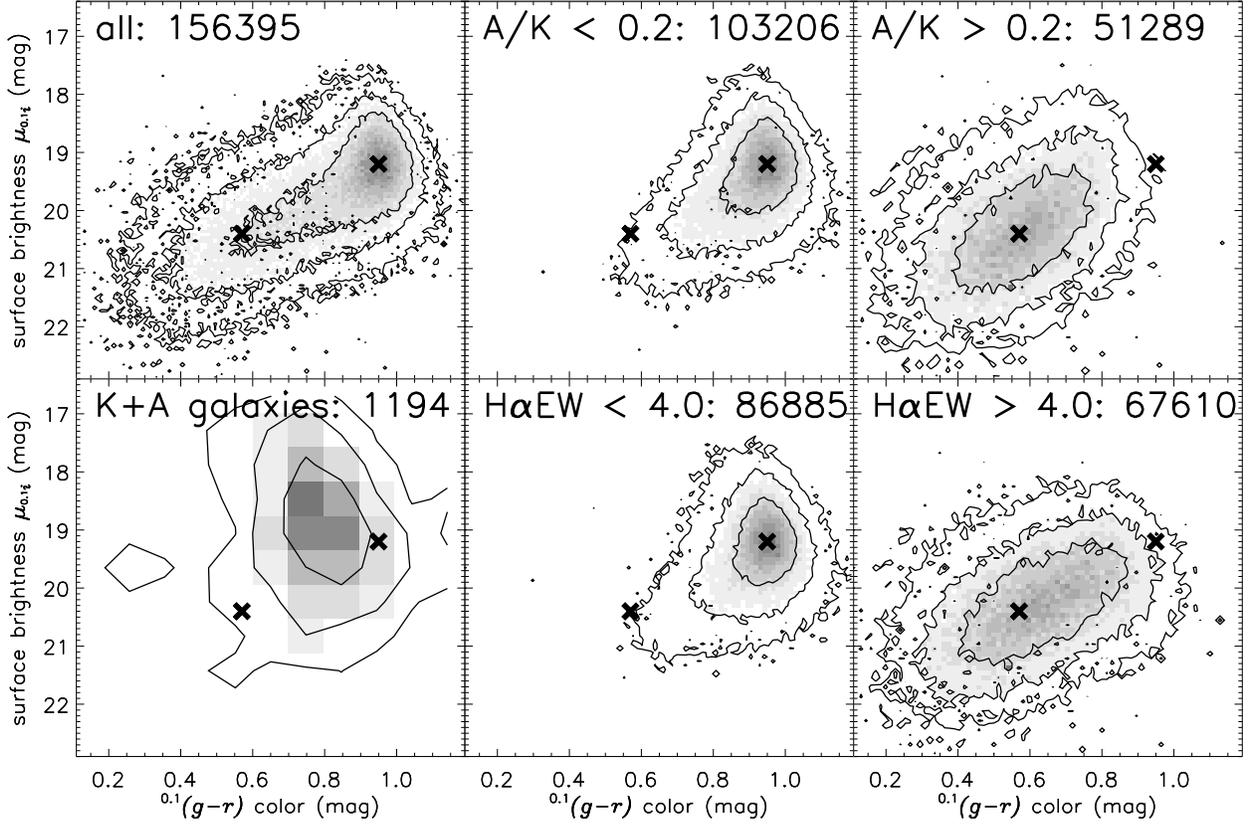}
\caption{Similar to Figure~\ref{fig:manyd_gmr_n} but showing the
distribution of luminosity density in galaxies in the sample in the
color--surface-brightness plane.  For each galaxy, the surface
brightness $\mu_{^{0.1}i}$ is the mean surface brightness inside the
Petrosian $R_{50}$ radius of the best-fitting S\'ersic radial profile
in the $^{0.1}i$ band.  In this panel, again, galaxies also separate
into disk-dominated (blue, low $\mu$) and bulge-dominated (red, high
$\mu$) populations, marked by crosses.  Galaxies selected to have low
$\Aamp /\Kamp$ or low \Halpha EW appear to be high in surface
brightness (bulge-dominated); galaxies selected to have high $\Aamp
/\Kamp$ or high \Halpha EW appear to be lower in surface brightness
(disk-dominated).  The K+A galaxies have the colors of disk-dominated
galaxies but many have central surface brightnesses as high as, or
even higher than, bulge-dominated galaxies.\label{fig:manyd_gmr_mu}}
\end{figure}

Figure~\ref{fig:manyd_gmr_mu} is similar to
Figure~\ref{fig:manyd_gmr_n} but shows the distribution of objects in
the color-surface-brightness plane. The population of objects in this
plane is not clearly bimodal, however the bulge-dominated galaxies
form a strong peak at high surface brightnesses and red colors. Just
like Figure~\ref{fig:manyd_gmr_n}, this shows that galaxies with high
A/K or high \Halpha EW appear to be disk-dominated and galaxies with
low A/K or low \Halpha EW appear to be bulge-dominated.  The A+K
population shows even higher surface brightnesses than the
bulge-dominated galaxies, suggesting that as they age, the A+K
galaxies will fade and redden into the bulge-dominated part of the
diagram.

Three-color images made from the SDSS imaging were inspected by eye
for a randomly selected sample of 160 of the K+A galaxies in the
sample.  The vast majority of them ($\sim90$~percent) appear bulge
dominated. The $\sim10$~percent that appear to be disk dominated do
also show significant bulges.  About $30$~percent either show tidal
features or lie extremely close on the sky to nearby neighbors,
indicating possible past or current interactions.

In Table~\ref{tab:odensity}, the mean overdensities
$\left<\delta_1\right>$ and $\left<\delta_8\right>$ on the $1\,h^{-1}$
and $8\,h^{-1}$~Mpc scales are given for the different galaxy
populations mentioned above.  Error bars on these averages were
computed by splitting the sample into three disjoint sky regions.
This technique produces conservative error estimates because it
effectively includes contributions from variations in calibration and
large-scale structure.  The numbers of galaxies used for mean
overdensities are smaller than used in the Figures because
overdensities have only been calculated on an unbiased subsample of
the sample we are using.  Table~\ref{tab:odensity} shows that K+A
galaxies, on average, live in the lower-density regions more typical
of spiral galaxies, and not the higher-density regions typical of
bulge-dominated galaxies.

\section{Discussion}

Two star-formation indicators, \Halpha EW and $\Aamp /\Kamp$, with two
different timescales, $\sim 10^7$~yr and $\sim 10^9$~yr, have been
measured in a sample of 156,395 SDSS (optically selected) galaxies
in the redshift range $0.05<z<0.20$.  The two star-formation
indicators are strongly correlated, as expected. Comparison with
models shows that the average levels of star formation in these
galaxies is below the average over cosmic time.  This fits in with
many observations of more distant galaxies that have implied a rapid
decrease in the global comoving star formation rate with cosmic time
to the present day \citep[][and references therein]{hogg01sfr}.

\begin{figure}
\plotone{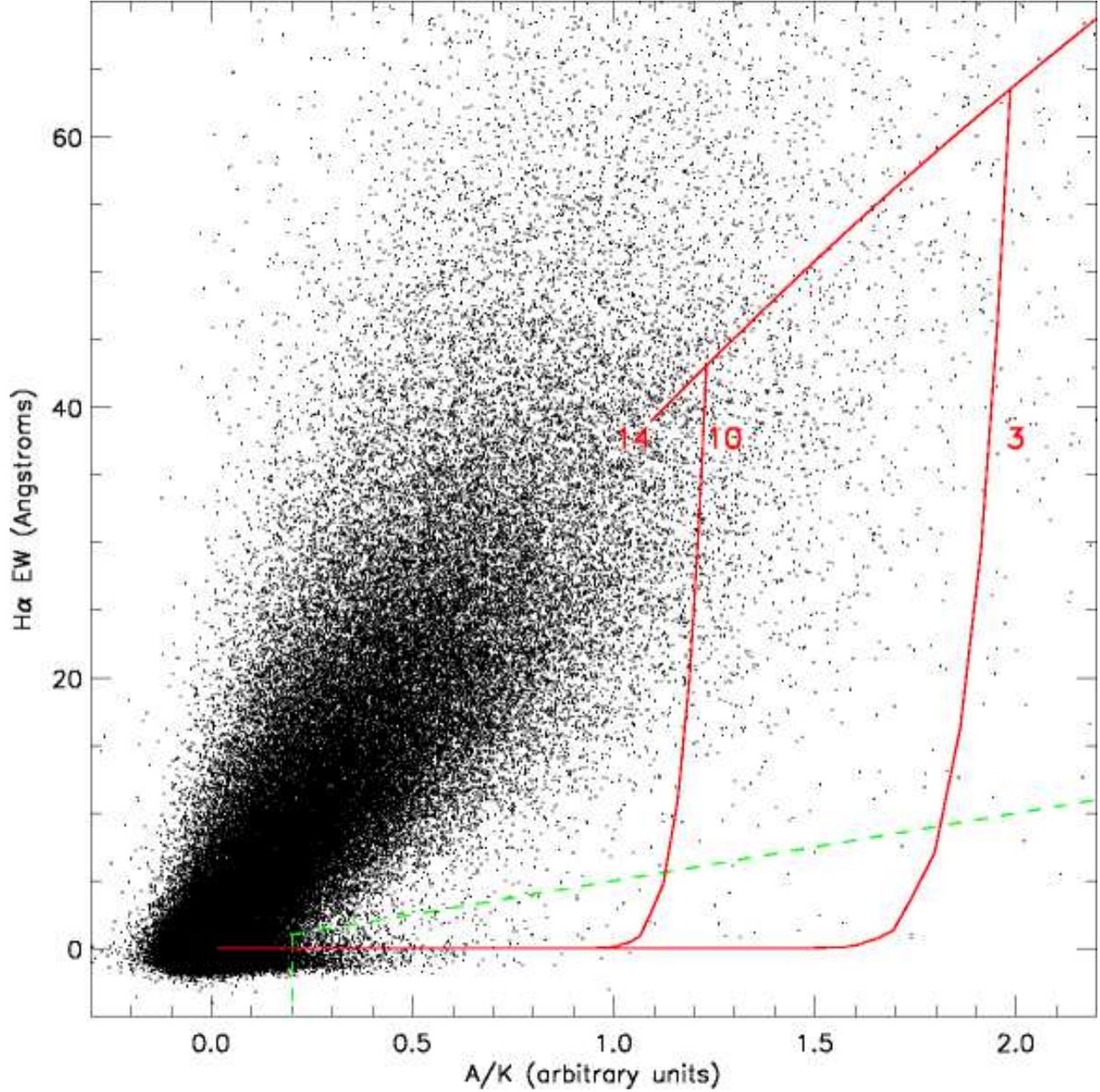}
\caption{The distribution of $\Aamp /\Kamp$ vs \Halpha\ for all the
galaxies in this sample.  Typical errors in $\Aamp /\Kamp$ are $\sim
0.03$ to $0.1$.  Shown in red are model tracks for the 3 models shown
in Figure~\ref{fig:Poster_timemodels}.  The K+A galaxy population was
selected with the cuts shown in green.\label{fig:Poster_AE_vs_Halpha}}
\end{figure}

A clearly separated, distinct population of ``K+A galaxies'' is found,
with very weak or no \Halpha EW, but significant $\Aamp /\Kamp$.  The
star-formation indicators \Halpha EW and $\Aamp /\Kamp$ have very
different timescales over which, in effect, they average recent star
formation rates.  Galaxies that abruptly (ie, on timescales
$<10^9$~yr) shut off star formation will fade rapidly in \Halpha EW,
but slowly in $\Aamp /\Kamp$.  This timescale mismatch explains the
presence of the distinctly visible K+A population shown in
\figurename~\ref{fig:Poster_AE_vs_Halpha}; we expect these galaxies to
indicate abruptly terminated star formation.

Bulge-dominated and disk-dominated galaxies can be separated by their
colors, concentrations (S\'ersic indices), and surface brightnesses.
By these measures, most of the K+A galaxies appear to be bluer, higher
surface-brightness counterparts of bulge-dominated galaxies.  The
offsets in color and surface brightness are consistent with the fading
and redenning expected in the evolution of stellar populations.  Taken
on face value, this is evidence that K+A galaxies are plausibly the
post-starburst progenitors of typical bulge-dominated galaxies, as
might be expected in a scenario in which disk-dominated galaxies, for
instance, merge to form bulge-dominated galaxies.

Interestingly, if that conclusion is correct, then these K+A galaxies
are the progenitors of the bulge-dominated galaxies of the field, not
of rich clusters, because the K+A galaxies live in the typical
overdensities of spirals, not ellipticals.  Past work on K+A galaxies
has implied that they form a special cluster population
\citep[eg,][]{dressler83a}, but this is presumably a consequence of
the selection in those studies.

\begin{figure}
\plotone{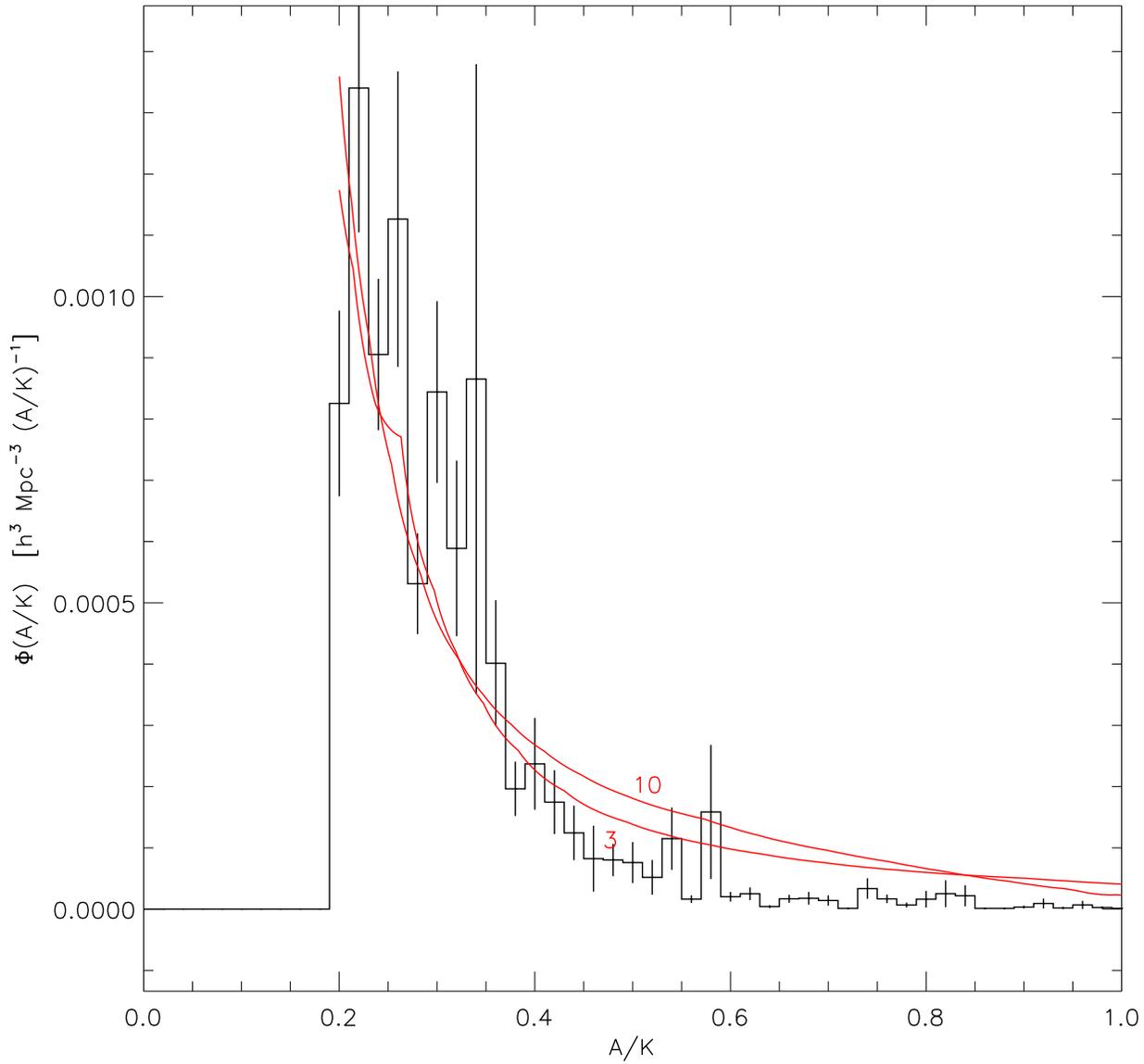}
\caption{The distribution of $\Aamp /\Kamp$, weighted by
$1/V_\mathrm{max}$, to make a ``ratio function''.  Overplotted are
curves proportional to $|\mathrm{d}t/\mathrm{d}(\Aamp /\Kamp)|$ for
the the ``10'' and ``3'' Gyr models of
Figure~\ref{fig:Poster_timemodels}.  These curves are equivalent to
``steady-state'' predictions made under the assumptions that (a) all
galaxies have the same star-formation history, and (b) galaxies are
passing through the A+K phase at a rate that varies slowly on Gyr
timescales.  The plotted uncertainties are simply the
$1/V_\mathrm{max}$ weights added in quadrature (divided by the bin
width).\label{fig:Poster_aehist}}
\end{figure}

In order to put the steady-state predictions onto
Figure~\ref{fig:Poster_aehist}, it was necessary to multiply the time
spent at each value of $\Aamp /\Kamp$, or
$|\mathrm{d}t/\mathrm{d}(\Aamp /\Kamp)|$ by a comoving rate density to
match the total abundance.  This rate density is 1.8 or 1.0 times
$10^{-4}\,h^3~\mathrm{Mpc^{-3}\,Gyr^{-1}}$ for the ``3'' or ``10''
model respectively.  In other words, the comoving K+A galaxy creation
rate density is on this order at redshifts $z\sim 0.1$.  This suggests
that of order 1~percent of galaxies are currently K+A, and of order
10~percent have been through a K+A phase since redshift unity.

The A and K spectra used for this study will be made available with
the electronically published version of this article.

\acknowledgements We thank Douglas Finkbeiner, Michel Fioc, Robert
Lupton, Bob Nichol, Jim Peebles, Don Schneider, David Spergel, Michael
Strauss, and Christy Tremonti for useful ideas, conversations, and
software.  This research made use of the NASA Astrophysics Data
System.  ADQ, DWH, and MRB are partially supported by NASA (grant
NAG5-11669) and NSF (grant PHY-0101738). DJE is supported by NSF
(grant AST-0098577) and by an Alfred P. Sloan Research Fellowship.

Funding for the creation and distribution of the SDSS Archive has been
provided by the Alfred P. Sloan Foundation, the Participating
Institutions, the National Aeronautics and Space Administration, the
National Science Foundation, the U.S. Department of Energy, the
Japanese Monbukagakusho, and the Max Planck Society. The SDSS Web site
is http://www.sdss.org/.

The SDSS is managed by the Astrophysical Research Consortium for the
Participating Institutions. The Participating Institutions are The
University of Chicago, Fermilab, the Institute for Advanced Study, the
Japan Participation Group, The Johns Hopkins University, Los Alamos
National Laboratory, the Max-Planck-Institute for Astronomy, the
Max-Planck-Institute for Astrophysics, New Mexico State University,
University of Pittsburgh, Princeton University, the United States
Naval Observatory, and the University of Washington.

\bibliographystyle{apj}
\bibliography{apj-jour,ccpp}

\clearpage
\begin{figure}
\plotone{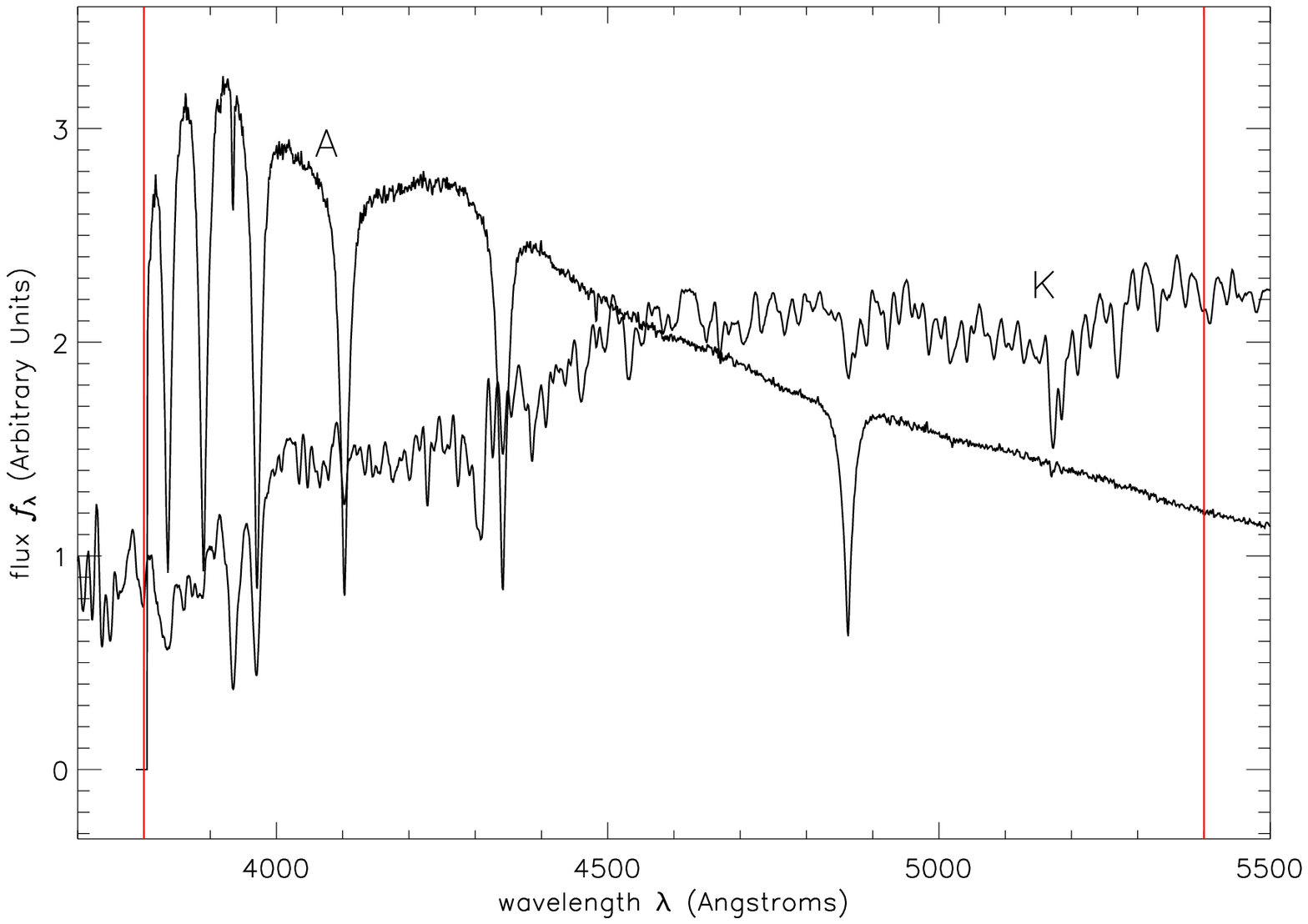}
\caption{The normalized model A-star spectrum (``A'') and average old
stellar population spectrum (``K'') used in the K+A fits.  The
vertical lines indicate the region over which the fit was
performed. As stated in the text, the spectra have equal areas inside
the fitting region.\label{fig:modelplots}}
\end{figure}

\clearpage
\begin{deluxetable}{lcccc}
\tablenum{1}
\tablewidth{0pc}
\tablecaption{average overdensities around galaxy subsamples
\label{tab:odensity}}
\tablehead{
\colhead{subsample} &
\colhead{$N$} &
\colhead{$j_{^{0.1}i}/j_\mathrm{total}$}\tablenotemark{a} &
\colhead{$\left<\delta_1\right>$} &
\colhead{$\left<\delta_8\right>$}
}
\startdata
all & 125255 & $1.000$ & $19.8 \pm  1.7$ & $1.86 \pm 0.19$ \\
disk-dominated ($n<2.0$) & 49615 & $0.433 \pm 0.007$ & $13.7 \pm  1.0$ & $1.55 \pm 0.19$ \\
bulge-dominated ($n>2.0$) & 75640 & $0.567 \pm 0.007$ & $24.5 \pm  2.2$ & $2.10 \pm 0.18$ \\
K+A galaxies (see text) & 1143 & $0.009 \pm 0.001$ & $15.0 \pm  2.3$ & $1.64 \pm 0.20$ \\

\enddata
\tablenotetext{a}{Fraction of the total galaxy luminosity density in
the $^{0.1}i$ band (\ie, total of $L_{^{0.1}i}/V_\mathrm{max}$)
contributed by each subsample.}
\end{deluxetable}

\end{document}